\documentclass[12pt]{article}
\usepackage[utf8]{inputenc}
\usepackage[T1]{fontenc}
\usepackage{amsmath,amsfonts,amssymb,amsthm}
\usepackage{mathtools}
\usepackage{geometry}
\usepackage{natbib}
\usepackage{graphicx}
\usepackage{booktabs}
\usepackage{longtable}
\usepackage{algorithm}
\usepackage{algorithmic}
\usepackage{hyperref}
\usepackage{xcolor}
\usepackage{setspace}
\usepackage{enumitem}
\usepackage{bm}
\usepackage{float}
\usepackage[title,titletoc]{appendix}
\newtheorem{theorem}{Theorem}[section]

\title{\textbf{Scalable and Communication-Efficient Varying Coefficient Mixed-Effects Models: Methodology, Theory, and Applications}}

\author{
Lida Chalangar Jalili Dehkharghani$^{1}$  and Li-Hsiang Lin$^{1*}$\\
  $^{1}$Department of Mathematics and Statistics,\\
  Georgia State University, Atlanta, GA, 30303, USA\\
}

\date{\today}

\begin{document}
\thispagestyle{empty} 
\maketitle

\begin{abstract}
Human migration exhibits complex spatiotemporal dependence driven by environmental and socioeconomic forces. Modeling such patterns at scale requires methods that accommodate many random effects while remaining feasible when raw data or large design matrices cannot be freely shared across distributed nodes. We develop a communication-efficient inference framework for Varying Coefficient Mixed Models (VCMMs) with flexible mean structures and large correlated random-effect components. Using a Bayesian hierarchical representation of penalized splines, we derive sufficient statistics that preserve each node's likelihood contribution and recover the estimator from the full data under unrestricted communication. Under communication constraints, these statistics support a one-step communication-efficient estimator with first-order efficiency. An SVD-enhanced implementation stabilizes large or ill-conditioned random-effect covariance operators. Theory establishes likelihood preservation, convergence, asymptotic efficiency, and finite-sample concentration. Simulations and U.S. migration-flow data demonstrate accuracy, scalability, and recovery of dynamic spatial patterns.
\end{abstract}

\textbf{Keywords:} Big Data Computation, Varying Coefficient Models, Sufficient Statistics, SVD Decomposition, Random Effects, Distributed Computing\\
\\
* Email of Corresponding Author is lhlin@gsu.edu

\newpage
\setcounter{page}{1} 
\section{Introduction}
\doublespacing

Analysis of human migration is central to understanding how population
redistribution reshapes urban resilience, labor-market adjustment,
infrastructure demand, economic opportunity, climate adaptation, and long-term
community vulnerability \citep{paglino2024estimating}. Recent advances in data
collection and computing have made it possible to record migration patterns at
increasingly large spatial and temporal scales. For example, at the Commuting
Zone (CZ) level, monthly movement records among \(154\) origin CZs and \(154\)
destination CZs from 2000 through 2020 yield
\(154\times154\times252=5{,}976{,}432\) CZ-to-CZ observations. Large-scale
migration records therefore contain millions of structured origin--destination
observations with hierarchical dependence, nonlinear time-varying patterns, and
spatial correlation. This scale motivates efficient statistical methods for
modeling migration flows while capturing time-varying covariate effects, origin
push effects, destination pull effects, and dependence among regional random
effects. Furthermore, administrative boundaries, privacy restrictions, and
computing-node limitations can make centralized pooling infeasible. These
challenges require methods that capture dependence, accommodate nonlinear
spatial--temporal patterns, scale to large data, and remain
communication-efficient.

A central object of migration analysis is the \textit{Origin--Destination (OD)
flow}, which records migration flows between regions over time
\citep{fields1979place,gurak1992migration}. Modern OD data pose several
challenges for existing statistical methods. First, OD flows depend on evolving
climate, economic, and social conditions, so varying-coefficient models are
needed to allow regression effects to change smoothly over time, space, or
context
\citep{hastie1993varying,hoover1998nonparametric,moore2020bayesian,
lu2009smoothing,cai2021efficient,chen2025exact}. Second, OD flows are
correlated through shared origins, shared destinations, regional environments,
and unobserved factors, which motivates random-effect formulations
\citep{wang2008random,li2010varying,chen2011penalized,morris2006wavelet}.
Together, these features lead naturally to Varying Coefficient Mixed Models
(VCMMs), which integrate smooth time-varying effects with random effects
\citep{tutz2003generalized,zhang2015estimation,li2020multilevel,
hung2022varying}. Existing VCMM methods, however, are usually developed for
centralized analysis and often rely on low-dimensional, sparse,
block-diagonal, separable, or pre-specified covariance assumptions. Such
restrictions are limiting for OD data, where origins, destinations, and
origin--destination interactions may induce dense, large-scale random-effect
covariance patterns. Thus, communication-efficient VCMM inference is called for
to simultaneously handle smooth varying effects, large random-effect components
with dense covariance dependence, and distributed computation.

To solve these problems, this paper develops a scalable and
communication-efficient inference framework for VCMMs with large-scale random
effects. We build on spline-based varying-coefficient estimation
\citep{eilers1996flexible,ruppert2003semiparametric,wood2017generalized} and
use its Bayesian hierarchical interpretation to derive likelihood-preserving
sufficient statistics. Although related Bayesian and penalized
varying-coefficient formulations have been studied
\citep{hodges2001counting,ruppert2003semiparametric,franco2019unified,
chen2025exact}, extensions to settings involving large dense random-effect
components, large-scale data, and communication constraints remain limited.

In distributed VCMM estimation, the central communication problem is to conduct
likelihood-based inference when each node can transmit only limited information
rather than raw data, spline-expanded design matrices, or random-effect design
matrices. This challenge is severe when the random-effect component is large,
densely correlated, and distributed across computational nodes, since
conventional mixed-model solvers often rely on sparse Cholesky factorization or
low-dimensional random-effect structures
\citep{lindstrom1988newton,muller2013model,bates2015fitting}. We address this problem by
replacing raw data and full design matrices with finite-dimensional sufficient
statistics computed locally and aggregated centrally. These summaries preserve
the mixed-model likelihood through the random effects and covariance operator.
The proposed algorithm allows the sufficient-statistics update to be iterated
multiple times, in which case it recovers the original full-data benchmark
without transmitting the original full data. It also allows the update to be
performed only once, yielding a one-step communication-efficient estimator with
first-order guarantees. When the aggregated random-effect block is large or
ill-conditioned, the proposed implementation is stabilized using large-scale
SVD techniques \citep{meng2014lsrn,qiu2024Rspectra}. Together, these results
establish likelihood preservation, numerical stability, asymptotic efficiency,
and communication-aware scalability for VCMM estimation.

Our framework differs from existing distributed estimation methods.
Divide-and-conquer methods \citep{zhang2015divide,chen2014split}, distributed
kernel regression \citep{wang2017efficient}, and communication-efficient sparse
learning
\citep{jordan2019communication,lee2017communication,fan2023communication}
mainly target independent observations and do not directly address
likelihood-preserving distributed inference for VCMMs with large-scale
dependence from dense random effects. Recent distributed varying-coefficient or
mixed-model methods address only part of the structure
\citep{huang2019distributed,guhaniyogi2022distribute,luo2022dlmm}. In
contrast, our method handles smooth varying coefficients, dense random-effect
dependence, and communication constraints in a likelihood-based framework.
In applications, this framework enables, to our knowledge, one of the largest
migration analyses jointly estimating origin push and destination pull random
effects with flexible dependence structure.

Although motivated by migration studies, the proposed framework applies broadly
to large structured and distributed data in climate science, biomedicine,
neuroimaging, and sensor networks. The remainder of the paper is organized as
follows. Section~\ref{sec:model} formulates the VCMM under a communication
constraint and develops the multi-round sufficient-statistics estimator,
including likelihood preservation, convergence, and SVD-stabilized
approximation. Section~\ref{sec:onestep_css} introduces the single-round
communication-efficient estimator and its theoretical guarantees.
Sections~\ref{sec:numericalstudies} and~\ref{sec:realdata} present simulations
and the migration application. Section~\ref{sec:conclusion} concludes, and the
supplemental material contains proofs, algorithms, and additional numerical
results.

\section{Problem Formulation and Sufficient Statistics for VCMMs}
\label{sec:model}
\label{sec:problemformulation}

We consider migration flow data observed across \(I\) origins and \(J\)
destinations over time \(t=1,\ldots,T\), so the total sample size is
\(N=IJT\). Let \(y_{ijt}\) denote the migration count from origin \(i\) to
destination \(j\) at time \(t\), with fixed-effect covariates
\(\mathbf x_{ijt}=(x_{1,ijt},\ldots,x_{p,ijt})^\top\in\mathbb R^p\) and
random-effect covariates \(\mathbf z_{ijt}\in\mathbb R^q\). We formulate a
general Varying Coefficient Mixed Model (VCMM) by allowing the conditional
distribution of \(y_{ijt}\) given the random effects to belong to an
exponential-family or quasi-likelihood class with conditional mean
\(\mu_{ijt}=E(y_{ijt}\mid\boldsymbol\alpha)\) and linear predictor
\begin{equation}
\label{eq:vcmm}
\eta_{ijt}
=
g(\mu_{ijt})
=
\beta_0(\mathbf h_{ijt})
+
\sum_{k=1}^{p}x_{k,ijt}\beta_k(\mathbf h_{ijt})
+
\mathbf z_{ijt}^{\top}\boldsymbol\alpha ,
\end{equation}
where \(g(\cdot)\) is a known link function and \(\mathbf h_{ijt}\) can contain
temporal, origin, destination, socioeconomic, or geographic information. For
migration flow counts, an important case is the log-linear VCMM with
$
\log(\mu_{ijt})
=
\beta_0(\mathbf h_{ijt})
+
\sum_{k=1}^{p}x_{k,ijt}\beta_k(\mathbf h_{ijt})
+
\mathbf z_{ijt}^{\top}\boldsymbol\alpha ,
$
which includes Poisson, quasi-Poisson, and negative-binomial count models. The
random effects satisfy
\(\boldsymbol\alpha\sim N(\mathbf 0,\boldsymbol\Sigma_\alpha)\), where
\(\dim(\boldsymbol\alpha)=q\) may be large and
\(\boldsymbol\Sigma_\alpha\) may encode dense dependence among origin effects,
destination effects, origin--destination interactions, temporal heterogeneity,
or their combinations. Throughout the paper, we use ``large random-effect
components'' to refer to settings where \(q\) is large and
\(\boldsymbol\Sigma_\alpha\) may be dense or weakly structured. The normal VCMM is obtained as a special case when \(g(\mu)=\mu\) with
\(\epsilon_{ijt}\sim N(0,\sigma_\epsilon^2)\).

For the VCMM in~\eqref{eq:vcmm}, large-scale applications require estimation
when raw observations, fixed-effect model matrices, and random-effect design
matrices are distributed across \(K\) computational nodes. Directly transmitting
these objects can be infeasible, especially when the spline dimension and the
random-effect dimension \(q\) are large. We therefore formulate distributed
likelihood-based estimation under a communication constraint. Let \(\mathbf y\)
denote the response vector obtained by stacking all observations \(y_{ijt}\) in
\eqref{eq:vcmm}, and let
\[
\boldsymbol\beta(\mathbf h)
=
\{\beta_0(\mathbf h),\beta_1(\mathbf h),\ldots,\beta_p(\mathbf h)\}^{\top}
\]
collect the varying coefficient functions. Then, our estimation problem can be
written as
\begin{equation}
\label{eq:contrainedoptimization}
\max_{\boldsymbol\beta(\mathbf h),\,\boldsymbol\alpha,\,\boldsymbol\eta}
\  \ell_{\mathrm{joint}}
\{\boldsymbol\beta(\mathbf h),\boldsymbol\alpha,\boldsymbol\eta\}
\quad
\text{subject to}
\quad
\text{total communication}\le cK d_{\Gamma},
\end{equation}
where $\ell_{\mathrm{joint}}
\{\boldsymbol\beta(\mathbf h),\boldsymbol\alpha,\boldsymbol\eta\}$ is the joint log likelihood of unknown parameters/functions $\boldsymbol\beta(\mathbf h)$, $\boldsymbol\alpha$, and $\boldsymbol\eta = (\boldsymbol\Sigma_\alpha,\sigma_\epsilon^2)$, and \(d_{\Gamma}\) denotes the dimension of the summary vector transmitted by
each node, and \(c\) denotes the number of communication rounds, where one
communication round means that all \(K\) nodes transmit their required
summaries to the central server once.

\noindent The goal is therefore to replace raw-data and full-design-matrix transmission
with local sufficient summaries while preserving the likelihood information
needed to estimate the varying coefficients, random effects, and variance
components.

\subsection{Iterative Sufficient-Statistics Estimation for VCMMs}
\label{subsec:sufficient_statistics_estimation}

The communication constraint in~\eqref{eq:contrainedoptimization} motivates
replacing raw observations and full design matrices by local summaries. We
first develop an iterative sufficient-statistics estimator for the regime where
multiple communication or refinement rounds are allowed. The main idea is to
replace each node's contribution to the current estimating objective by a
finite-dimensional local quadratic summary. These local summaries are then
aggregated at the central server, which updates the varying-coefficient
parameters, random effects, and nuisance parameters without requiring raw data
or full design matrices. This estimator provides the full-data iterative
benchmark under the proposed communication framework and identifies the summary
structure used again by the one-step estimator in
Section~\ref{sec:onestep_css}.

We represent the varying coefficient functions in~\eqref{eq:vcmm} using
tensor-product P-splines \citep{eilers1996flexible,ruppert2003semiparametric}.
Let \(m_\beta=(p+1)Q\) be the total number of spline coefficients after basis
expansion, and let
\(\tilde{\boldsymbol\beta}\in\mathbb R^{m_\beta}\) collect all spline
coefficients. After stacking all observations, the linear predictor can be
written as
\begin{equation}
\label{eq:linear_predictor_matrix}
\boldsymbol\xi
=
\tilde{\mathbf X}\tilde{\boldsymbol\beta}
+
\mathbf Z\boldsymbol\alpha,
\end{equation}
where \(\tilde{\mathbf X}\in\mathbb R^{N\times m_\beta}\) is the
spline-expanded fixed-effect design matrix and
\(\mathbf Z\in\mathbb R^{N\times q}\) is the random-effect design matrix. The
Bayesian P-spline formulation assigns the Gaussian penalty prior
$\pi(\tilde{\boldsymbol\beta}\mid\lambda^*)
\propto
\exp\{-\tilde{\boldsymbol\beta}^{\top}
\mathbf P_{\lambda^*}\tilde{\boldsymbol\beta}/2\},$
where \(\mathbf P_{\lambda^*}\) is the P-spline penalty matrix, and assigns the
Gaussian prior
$\boldsymbol\alpha\mid\boldsymbol\Sigma_\alpha
\sim
N(\mathbf 0,\boldsymbol\Sigma_\alpha)$
to the random effects.

For the general exponential-family or log-linear VCMM, the penalized likelihood
is not necessarily quadratic in
\((\tilde{\boldsymbol\beta},\boldsymbol\alpha)\). Therefore, we use an
iterative working-quadratic representation. At iteration \(r\), the current
estimates
$(\widehat{\tilde{\boldsymbol\beta}}^{(r)},
\widehat{\boldsymbol\alpha}^{(r)},
\widehat{\boldsymbol\eta}^{(r)})$
determine the current linear predictor
$
\widehat{\boldsymbol\xi}^{(r)}
=
\tilde{\mathbf X}\widehat{\tilde{\boldsymbol\beta}}^{(r)}
+
\mathbf Z\widehat{\boldsymbol\alpha}^{(r)},
$
the current conditional mean
$
\widehat{\boldsymbol\mu}^{(r)}
=
E(\mathbf y\mid \widehat{\boldsymbol\xi}^{(r)}),
$
and a local quadratic approximation to the negative log likelihood. This
approximation can be written in weighted least-squares form as
\begin{equation}
\label{eq:working_quadratic_objective}
\mathcal L^{(r)}
(\tilde{\boldsymbol\beta},\boldsymbol\alpha;\boldsymbol\eta)
=
\frac{1}{2}
\left\|
\mathbf W^{(r)1/2}
\left\{
\mathbf y^{*(r)}
-
\tilde{\mathbf X}\tilde{\boldsymbol\beta}
-
\mathbf Z\boldsymbol\alpha
\right\}
\right\|^2
+
\frac{1}{2}
\tilde{\boldsymbol\beta}^{\top}
\mathbf P_{\lambda^*}
\tilde{\boldsymbol\beta}
+
\frac{1}{2}
\boldsymbol\alpha^{\top}
\boldsymbol\Sigma_\alpha^{-1}
\boldsymbol\alpha ,
\end{equation}
where \(\mathbf y^{*(r)}\) is the working response and
\(\mathbf W^{(r)}\) is the working weight matrix. For an IRLS or
Fisher-scoring approximation,
\begin{equation}
\label{eq:working_response_weight}
\mathbf y^{*(r)}
=
\widehat{\boldsymbol\xi}^{(r)}
+
\mathbf D^{(r)-1}
\{\mathbf y-\widehat{\boldsymbol\mu}^{(r)}\},
\qquad
\mathbf W^{(r)}
=
\mathbf D^{(r)}
\mathbf V^{(r)-1}
\mathbf D^{(r)},
\end{equation}
where
\(\mathbf D^{(r)}=\partial\boldsymbol\mu/
\partial\boldsymbol\xi^\top\) is evaluated at the current estimates and
\(\mathbf V^{(r)}\) is the conditional variance matrix or a working
approximation to it. For example, for a Poisson log-linear VCMM,
$
\widehat\mu_i^{(r)}=\exp(\widehat\xi_i^{(r)})$,
$y_i^{*(r)}
=
\widehat\xi_i^{(r)}
+
\frac{y_i-\widehat\mu_i^{(r)}}{\widehat\mu_i^{(r)}}$,
$w_i^{(r)}
=
\widehat\mu_i^{(r)}.
$
For a negative-binomial log-linear VCMM with
$\mathrm{Var}(y_i\mid\boldsymbol\alpha)
=
\mu_i+\kappa\mu_i^2,$
the corresponding working weight is
$
w_i^{(r)}
=
\frac{\widehat\mu_i^{(r)}}{1+\kappa\widehat\mu_i^{(r)}} .
$
Thus, in log-linear and generalized VCMMs, both the working response and the
working weights depend on the current estimates and must be recomputed across
refinement rounds.

To express~\eqref{eq:working_quadratic_objective} under distributed storage,
let
$
\mathcal D_s=(\mathbf y_s,\tilde{\mathbf X}_s,\mathbf Z_s),
\qquad
s=1,\ldots,K,
$
denote the data partition on node \(s\). At iteration \(r\), node \(s\)
constructs the local working response \(\mathbf y_s^{*(r)}\) and working
weight matrix \(\mathbf W_s^{(r)}\). Define the round-\(r\) local summary
\[
\boldsymbol\Gamma_s^{(r)}
=
(a_s^{(r)},\mathbf b_s^{(r)},\mathbf C_s^{(r)},
\mathbf d_s^{(r)},\mathbf B_s^{(r)},\mathbf H_s^{(r)}),
\]
where
\[
a_s^{(r)}
=
\mathbf y_s^{*(r)\top}
\mathbf W_s^{(r)}
\mathbf y_s^{*(r)},
\qquad
\mathbf b_s^{(r)}
=
\tilde{\mathbf X}_s^\top
\mathbf W_s^{(r)}
\mathbf y_s^{*(r)},
\]
\[
\mathbf C_s^{(r)}
=
\tilde{\mathbf X}_s^\top
\mathbf W_s^{(r)}
\tilde{\mathbf X}_s,
\qquad
\mathbf d_s^{(r)}
=
\mathbf Z_s^\top
\mathbf W_s^{(r)}
\mathbf y_s^{*(r)},
\]
\[
\mathbf B_s^{(r)}
=
\tilde{\mathbf X}_s^\top
\mathbf W_s^{(r)}
\mathbf Z_s,
\qquad
\mathbf H_s^{(r)}
=
\mathbf Z_s^\top
\mathbf W_s^{(r)}
\mathbf Z_s .
\]
The central server aggregates the local summaries as
$
\boldsymbol\Gamma^{(r)}
=
(a^{(r)},\mathbf b^{(r)},\mathbf C^{(r)},
\mathbf d^{(r)},\mathbf B^{(r)},\mathbf H^{(r)})
=
\sum_{s=1}^{K}\boldsymbol\Gamma_s^{(r)}.
$
Expanding the squared term in
\eqref{eq:working_quadratic_objective} shows that the round-\(r\) working
objective depends on the distributed data only through
\(\boldsymbol\Gamma^{(r)}\). Specifically, the summary-based working objective
is
\begin{equation}
\label{eq:ss_summary_objective}
\begin{aligned}
\mathcal L_{\Gamma}^{(r)}
(\tilde{\boldsymbol\beta},\boldsymbol\alpha;\boldsymbol\eta)
&=
\frac{1}{2}
\left[
a^{(r)}
-2\tilde{\boldsymbol\beta}^{\top}\mathbf b^{(r)}
-2\boldsymbol\alpha^\top\mathbf d^{(r)}
+\tilde{\boldsymbol\beta}^{\top}\mathbf C^{(r)}\tilde{\boldsymbol\beta}
+2\tilde{\boldsymbol\beta}^{\top}\mathbf B^{(r)}\boldsymbol\alpha
+\boldsymbol\alpha^\top\mathbf H^{(r)}\boldsymbol\alpha
\right]  \\
&\quad
+
\frac{1}{2}\tilde{\boldsymbol\beta}^{\top}
\mathbf P_{\lambda^*}\tilde{\boldsymbol\beta}
+
\frac{1}{2}\boldsymbol\alpha^{\top}
\boldsymbol\Sigma_\alpha^{-1}\boldsymbol\alpha .
\end{aligned}
\end{equation}
The update of
\((\tilde{\boldsymbol\beta},\boldsymbol\alpha)\) is obtained by minimizing
\(\mathcal L_{\Gamma}^{(r)}
(\tilde{\boldsymbol\beta},\boldsymbol\alpha;\boldsymbol\eta)\)
with the current nuisance-parameter estimates. When nuisance parameters are
unknown, they are updated by a fixed rule
\begin{equation}
\label{eq:eta_update_rule}
\boldsymbol\eta^{(r+1)}
=
\mathcal M_{\eta}\{
\boldsymbol\Gamma^{(r)},
\tilde{\boldsymbol\beta}^{(r+1)},
\boldsymbol\alpha^{(r+1)}
\},
\end{equation}
where \(\mathcal M_{\eta}\) may correspond to an ML, REML,
quasi-likelihood, Fisher-scoring, or moment-based update
\citep{jiang2007linear}. The same working responses, working weights, update
rule, and initialization are used for both the distributed sufficient-statistics
algorithm and the corresponding centralized full-data algorithm. Therefore, the
distributed algorithm reproduces the centralized working-quadratic update at
each round while transmitting only local summaries rather than raw observations
or full design matrices. The following theorem formalizes this round-wise preservation property. The
detailed proof is given in Appendix A of the
Supplemental Material.

\begin{theorem}[Round-wise likelihood-preserving sufficient statistics for VCMM estimation]
\label{thm:sufficient}
Consider the working-quadratic objective
\eqref{eq:working_quadratic_objective} at iteration \(r\), and suppose that
each node constructs
\(\boldsymbol\Gamma_s^{(r)}\) using the same working response
\(\mathbf y_s^{*(r)}\) and working weight matrix \(\mathbf W_s^{(r)}\) as the
corresponding centralized algorithm. Then the aggregated summary
\(\boldsymbol\Gamma^{(r)}=\sum_{s=1}^K\boldsymbol\Gamma_s^{(r)}\) is
likelihood-preserving for the round-\(r\) working objective in the following
sense.

\begin{enumerate}
\item[\textnormal{(i)}]
For fixed nuisance parameters \(\boldsymbol\eta\), the minimizer of the
centralized working objective
\(\mathcal L^{(r)}
(\tilde{\boldsymbol\beta},\boldsymbol\alpha;\boldsymbol\eta)\)
in~\eqref{eq:working_quadratic_objective} is identical to the minimizer of the
summary-based working objective
\(\mathcal L_{\Gamma}^{(r)}
(\tilde{\boldsymbol\beta},\boldsymbol\alpha;\boldsymbol\eta)\)
in~\eqref{eq:ss_summary_objective}.

\item[\textnormal{(ii)}]
If further under conditions (A1) to (A4) in
Appendix A, the nuisance parameters are updated by the
fixed rule \(\mathcal M_\eta\) in~\eqref{eq:eta_update_rule}, and the same
initialization is used, then for any \(R_{\max}\le c\), the distributed
summary-based iterates
\[
\widehat{\boldsymbol\theta}_{\Gamma}^{(r)}
=
(\widehat{\tilde{\boldsymbol\beta}}_{\Gamma}^{(r)\top},
\widehat{\boldsymbol\alpha}_{\Gamma}^{(r)\top})^\top,
\qquad
r=1,\ldots,R_{\max},
\]
are identical to the iterates obtained by applying the same working-quadratic
algorithm to the original full data. Moreover,
\(\widehat{\boldsymbol\theta}_{\Gamma}^{(R_{\max})}\) is feasible for the
communication-constrained problem in~\eqref{eq:contrainedoptimization}.
\end{enumerate}
\end{theorem}

\noindent Theorem~\ref{thm:sufficient} shows that the distributed summaries
preserve the information needed for each centralized working-quadratic update.
After each node transmits
\(\boldsymbol\Gamma_s^{(r)}\), the raw observations, the
\(N\times m_\beta\) spline-expanded design matrix, and the
\(N\times q\) random-effect design matrix are no longer needed for the
round-\(r\) update. All central computations use aggregated matrices of sizes
\(m_\beta\times m_\beta\), \(m_\beta\times q\), and \(q\times q\). This
reduction is especially useful when \(q\) is large. If the aggregated
random-effect block is dense or ill-conditioned, the random-effect update can
be stabilized by a truncated or regularized SVD applied directly to the
aggregated block; see Appendix B.1.

Theorem~\ref{thm:sufficient} leads directly to the iterative
sufficient-statistics estimator in Algorithm~\ref{alg:ss}. At each refinement
round, the central server broadcasts the current estimates to the nodes. Each
node then constructs a local working response and working weight matrix,
computes its round-specific local summary, and transmits this summary to the
central server. The server aggregates the summaries and updates the spline
coefficients, random effects, and nuisance parameters using
\eqref{eq:ss_summary_objective} and \eqref{eq:eta_update_rule}. Because the
working responses and working weights generally depend on the current estimates
in log-linear and generalized VCMMs, the summaries
\(\boldsymbol\Gamma_s^{(r)}\) must be refreshed across communication rounds.

\begin{algorithm}[h!]
\caption{Iterative Sufficient-Statistics Estimation for VCMMs}
\label{alg:ss}
\begin{algorithmic}[1]
\STATE \textbf{Input:}
Data partitions \(\{\mathcal D_s\}_{s=1}^K\), penalty matrix
\(\mathbf P_{\lambda^*}\), initial estimates
\((\widehat{\tilde{\boldsymbol\beta}}^{(0)},
\widehat{\boldsymbol\alpha}^{(0)},
\widehat{\boldsymbol\eta}^{(0)})\), nuisance-parameter update rule
\(\mathcal M_\eta\), tolerance \(\varepsilon\), maximum number of rounds
\(R_{\max}\le c\), and optional SVD-stabilization parameters.

\FOR{\(r=0,1,\ldots,R_{\max}-1\)}

    \STATE Broadcast the current estimates
    \((\widehat{\tilde{\boldsymbol\beta}}^{(r)},
    \widehat{\boldsymbol\alpha}^{(r)},
    \widehat{\boldsymbol\eta}^{(r)})\)
    to all nodes.

    \FOR{each node \(s=1,\ldots,K\)}
        \STATE Construct the current working response
        \(\mathbf y_s^{*(r)}\) and working weight matrix
        \(\mathbf W_s^{(r)}\).
        \STATE Compute the round-\(r\) local summary
        \(\boldsymbol\Gamma_s^{(r)}
        =(a_s^{(r)},\mathbf b_s^{(r)},\mathbf C_s^{(r)},
        \mathbf d_s^{(r)},\mathbf B_s^{(r)},\mathbf H_s^{(r)})\).
        \STATE Transmit \(\boldsymbol\Gamma_s^{(r)}\) to the central server.
    \ENDFOR

    \STATE Aggregate the local summaries to obtain
    \[
    \boldsymbol\Gamma^{(r)}
    =
    \sum_{s=1}^{K}\boldsymbol\Gamma_s^{(r)}.
    \]

    \STATE Update
    \((\widehat{\tilde{\boldsymbol\beta}}^{(r+1)},
    \widehat{\boldsymbol\alpha}^{(r+1)})\)
    by minimizing the summary-based working objective in
    \eqref{eq:ss_summary_objective} with the current nuisance-parameter
    estimate \(\widehat{\boldsymbol\eta}^{(r)}\).

    \STATE If the aggregated random-effect block in the
    \(\boldsymbol\alpha\)-update is ill-conditioned, use the SVD-stabilized
    update described in Appendix B.1.

    \STATE Update
    \(\widehat{\boldsymbol\eta}^{(r+1)}\) using
    \[
    \mathcal M_\eta\{
    \boldsymbol\Gamma^{(r)},
    \widehat{\tilde{\boldsymbol\beta}}^{(r+1)},
    \widehat{\boldsymbol\alpha}^{(r+1)}
    \}.
    \]

    \STATE Stop if the relative change in
    \((\widehat{\tilde{\boldsymbol\beta}},
    \widehat{\boldsymbol\alpha},
    \widehat{\boldsymbol\eta})\)
    is below \(\varepsilon\).

\ENDFOR

\STATE \textbf{Output:}
\(\widehat{\tilde{\boldsymbol\beta}}\),
\(\widehat{\boldsymbol\alpha}\), and
\(\widehat{\boldsymbol\eta}\).
\end{algorithmic}
\end{algorithm}

\noindent Algorithm~\ref{alg:ss} separates communication and computation across
refinement rounds. In each round, every node transmits only its local summary,
while all parameter and nuisance-parameter updates are carried out at the
central server using the aggregated summaries. Since
\(\mathbf C_s^{(r)}\) and \(\mathbf H_s^{(r)}\) are symmetric, the
communication dimension per node in each round is
\begin{equation}
\label{eq:d_gamma}
d_\Gamma
=
1
+
m_\beta
+
\frac{m_\beta(m_\beta+1)}{2}
+
q
+
m_\beta q
+
\frac{q(q+1)}{2}.
\end{equation}
Therefore, \(R_{\max}\) refinement rounds require total communication
\(R_{\max}K d_\Gamma\), and the estimator is feasible under the communication
constraint in~\eqref{eq:contrainedoptimization} whenever \(R_{\max}\le c\).
Hence, the communication cost depends on the spline dimension \(m_\beta\), the
random-effect dimension \(q\), and the number of refinement rounds, but not on
the local sample size. Algorithm~\ref{alg:ss} therefore provides the iterative
sufficient-statistics estimator for the \(c>1\) regime. The next subsection
will consider the stricter \(c=1\) regime, where only a single
communication-efficient refinement is performed from a pilot estimator.

Note that the normal linear VCMM is a special case of this general framework. In this
case,
$
\mathbf y
=
\tilde{\mathbf X}\tilde{\boldsymbol\beta}
+
\mathbf Z\boldsymbol\alpha
+
\boldsymbol\epsilon$,
$\boldsymbol\epsilon\sim
N(\mathbf 0,\sigma_\epsilon^2\mathbf I_N),
$
and the working response and working weight matrix can be taken as
$\mathbf y^{*(r)}=\mathbf y$ and
$\mathbf W^{(r)}=\sigma_\epsilon^{-2}\mathbf I_N$.
Thus, the local summaries reduce to fixed cross-products and do not depend on
the current parameter estimates. Consequently, for the normal linear VCMM with
a fixed spline basis and homoscedastic Gaussian errors, one transmission of the
fixed sufficient summaries is enough to reproduce the full-data Gaussian
objective, and the multi-round communication scheme collapses to a
one-communication sufficient-statistics estimator followed by central
iterations. In contrast, for log-linear and other generalized VCMMs, the
working responses and working weights change with the current estimates, so the
round-specific summaries must be recomputed and retransmitted across
refinement rounds.

To provide numerical support for Algorithm~\ref{alg:ss}, we first conduct a
validation experiment in which the centralized full-data estimator can be
compared directly with the sufficient-statistics estimator. As reported in
Table 2 of Appendix F.1, the SS
estimator is nearly identical to the conventional centralized estimator, with
correlations exceeding \(0.998\) for the intercept, the varying coefficient,
and the variance component. The MSEs are also of the same order. These results
support the round-wise preservation property in
Theorem~\ref{thm:sufficient} and confirm that the proposed summaries retain the
information needed to reproduce the corresponding full-data iterative update.
Full simulation details are provided in Appendix F.1.

Theorem~\ref{thm:sufficient} establishes the role of Algorithm~\ref{alg:ss} in
the proposed communication framework. Under the \(c>1\) regime in
\eqref{eq:contrainedoptimization}, the proposed algorithm uses the likelihood
information contained in the aggregated round-specific summaries and refines
the parameter estimates until convergence. When the communication budget allows
only one communication round, we instead use the one-step estimator developed
in the next section.

\section{One-Step Communication-Efficient Estimation for VCMMs}
\label{sec:onestep_css}

We now consider the stricter case \(c=1\) in
\eqref{eq:contrainedoptimization}, where only one communication round is
allowed after a pilot estimator is available. Let
\(\boldsymbol\theta=(\tilde{\boldsymbol\beta}^{\top},
\boldsymbol\alpha^{\top})^{\top}\) collect the spline coefficients and random
effects, and let
\(\widehat{\boldsymbol\theta}_0
=
(\widehat{\tilde{\boldsymbol\beta}}_0^{\top},
\widehat{\boldsymbol\alpha}_0^{\top})^{\top}\)
denote a pilot estimator. In this paper, the pilot estimator is obtained from
the first node, which is also used as the reference node in the curvature
approximation. The corresponding pilot nuisance-parameter estimator is denoted
by
\(\widehat{\boldsymbol\eta}_0
=
(\widehat{\boldsymbol\Sigma}_{\alpha,0},
\widehat\sigma_{\epsilon,0}^2)\).
For generalized or log-linear VCMMs, the pilot estimators
\((\widehat{\boldsymbol\theta}_0,\widehat{\boldsymbol\eta}_0)\) determine the
pilot working responses and pilot working weights used to form the one-step
summary. For the normal linear special case, this corresponds to
\(\mathbf y_s^{*(0)}=\mathbf y_s\) and
\(\mathbf W_s^{(0)}=\widehat\sigma_{\epsilon,0}^{-2}\mathbf I_s\).

The main idea of the one-step estimator is to use the distributed data only
once to evaluate a global first-order correction at the pilot estimator, while
replacing the full aggregated curvature by a reference-node curvature
approximation. Thus, the method uses one round of communication to construct
the pilot-based global score, but it does not iterate the sufficient-statistics
updates to convergence and does not invert the full aggregated curvature matrix
as in the SS estimator.

Using the pilot-based working summaries
$$\boldsymbol\Gamma^{(0)}
=
(a^{(0)},\mathbf b^{(0)},\mathbf C^{(0)},
\mathbf d^{(0)},\mathbf B^{(0)},\mathbf H^{(0)})=
\sum_{s=1}^K
(a_s^{(0)},\mathbf b_s^{(0)},\mathbf C_s^{(0)},
\mathbf d_s^{(0)},\mathbf B_s^{(0)},\mathbf H_s^{(0)}),$$
the gradient of the pilot working objective at
\(\widehat{\boldsymbol\theta}_0\) is
\begin{equation}
\label{eq:onestep_gradient}
\mathbf g(\widehat{\boldsymbol\theta}_0;\widehat{\boldsymbol\eta}_0)
=
\begin{pmatrix}
\mathbf g_{\tilde{\boldsymbol\beta}}
(\widehat{\boldsymbol\theta}_0;\widehat{\boldsymbol\eta}_0)
\\[0.25em]
\mathbf g_{\boldsymbol\alpha}
(\widehat{\boldsymbol\theta}_0;\widehat{\boldsymbol\eta}_0)
\end{pmatrix},
\end{equation}
where
\begin{equation}
\label{eq:onestep_gradient_components}
\begin{aligned}
\mathbf g_{\tilde{\boldsymbol\beta}}
(\widehat{\boldsymbol\theta}_0;\widehat{\boldsymbol\eta}_0)
&=
\mathbf C^{(0)}\widehat{\tilde{\boldsymbol\beta}}_0
+
\mathbf B^{(0)}\widehat{\boldsymbol\alpha}_0
-
\mathbf b^{(0)}
+
\mathbf P_{\lambda^*}\widehat{\tilde{\boldsymbol\beta}}_0,
\\
\mathbf g_{\boldsymbol\alpha}
(\widehat{\boldsymbol\theta}_0;\widehat{\boldsymbol\eta}_0)
&=
\mathbf B^{(0)\top}\widehat{\tilde{\boldsymbol\beta}}_0
+
\mathbf H^{(0)}\widehat{\boldsymbol\alpha}_0
-
\mathbf d^{(0)}
+
\widehat{\boldsymbol\Sigma}_{\alpha,0}^{-1}
\widehat{\boldsymbol\alpha}_0 .
\end{aligned}
\end{equation}
These expressions use the aggregated pilot-based summaries only to compute the
global gradient at the pilot estimator. The dependence on
\(\widehat\sigma_{\epsilon,0}^2\) is absorbed through the pilot working weight
matrix \(\mathbf W^{(0)}\). In particular, for the normal linear special case,
\(\mathbf W^{(0)}=\widehat\sigma_{\epsilon,0}^{-2}\mathbf I_N\), so the
weighted summaries in \(\boldsymbol\Gamma^{(0)}\) are equivalent to multiplying
the usual Gaussian cross-products by
\(\widehat\sigma_{\epsilon,0}^{-2}\).

The curvature used in the one-step correction is not the full aggregated
curvature. Instead, we use a reference-node curvature approximation. Since the
pilot estimator is obtained from node \(1\), we take node \(1\) as the
reference node. Let \(n_1\) be its local sample size and set \(w_1=N/n_1\).
Define
\begin{equation}
\label{eq:onestep_reference_curvature}
\widetilde{\boldsymbol{\mathcal K}}
=
\begin{pmatrix}
w_1\mathbf C_1^{(0)}+\mathbf P_{\lambda^*}
&
w_1\mathbf B_1^{(0)}
\\[0.35em]
w_1\mathbf B_1^{(0)\top}
&
w_1\mathbf H_1^{(0)}
+
\widehat{\boldsymbol\Sigma}_{\alpha,0}^{-1}
\end{pmatrix},
\end{equation}
where
\(\mathbf C_1^{(0)}\), \(\mathbf B_1^{(0)}\), and
\(\mathbf H_1^{(0)}\) are the pilot-based working summaries computed on node
\(1\). The scaling factor \(w_1=N/n_1\) places the reference-node likelihood
curvature on the same order as the full-data working curvature, while the
penalty and prior curvature terms \(\mathbf P_{\lambda^*}\) and
\(\widehat{\boldsymbol\Sigma}_{\alpha,0}^{-1}\) are not scaled.

The one-step communication-efficient estimator is
\begin{equation}
\label{eq:onestep_update}
\widehat{\boldsymbol\theta}_{\mathrm{OS}}
=
\widehat{\boldsymbol\theta}_0
-
\widetilde{\boldsymbol{\mathcal K}}^{-1}
\mathbf g(\widehat{\boldsymbol\theta}_0;\widehat{\boldsymbol\eta}_0).
\end{equation}
Thus, the OS estimator is a single Newton-type correction from the pilot
estimator. The correction direction is determined by the global score in
\eqref{eq:onestep_gradient}--\eqref{eq:onestep_gradient_components}, while the
curvature adjustment is supplied by the reference-node matrix
\(\widetilde{\boldsymbol{\mathcal K}}^{-1}\). This is the key distinction from
the iterative SS estimator: OS uses one communication round and one correction,
whereas SS refreshes the working summaries across communication rounds and
updates to convergence.

Writing
\(\widehat{\boldsymbol\theta}_{\mathrm{OS}}
=
(\widehat{\tilde{\boldsymbol\beta}}_{\mathrm{OS}}^{\top},
\widehat{\boldsymbol\alpha}_{\mathrm{OS}}^{\top})^{\top}\),
the nuisance parameters are updated by the same rule used in the SS algorithm:
\begin{equation}
\label{eq:eta_onestep}
\widehat{\boldsymbol\eta}_{\mathrm{OS}}
=
\mathcal M_{\eta}
\{
\boldsymbol\Gamma^{(0)},
\widehat{\tilde{\boldsymbol\beta}}_{\mathrm{OS}},
\widehat{\boldsymbol\alpha}_{\mathrm{OS}}
\}.
\end{equation}
In particular,
\(\widehat{\boldsymbol\eta}_{\mathrm{OS}}\) contains the updated variance
component estimators
\(\widehat{\boldsymbol\Sigma}_{\alpha,\mathrm{OS}}\) and
\(\widehat\sigma_{\epsilon,\mathrm{OS}}^2\). For the normal linear special
case, if the unweighted Gaussian summaries are also used for the residual
variance update, then
\begin{equation}
\label{eq:sigma_onestep}
\widehat{\sigma}_{\epsilon,\mathrm{OS}}^2
=
\frac{1}{N}
\left[
a
-2\widehat{\tilde{\boldsymbol\beta}}_{\mathrm{OS}}^\top\mathbf b
+\widehat{\tilde{\boldsymbol\beta}}_{\mathrm{OS}}^\top
\mathbf C
\widehat{\tilde{\boldsymbol\beta}}_{\mathrm{OS}}
-2\widehat{\boldsymbol\alpha}_{\mathrm{OS}}^\top\mathbf d
+2\widehat{\tilde{\boldsymbol\beta}}_{\mathrm{OS}}^\top
\mathbf B
\widehat{\boldsymbol\alpha}_{\mathrm{OS}}
+\widehat{\boldsymbol\alpha}_{\mathrm{OS}}^\top
\mathbf H
\widehat{\boldsymbol\alpha}_{\mathrm{OS}}
\right],
\end{equation}
where \(a,\mathbf b,\mathbf C,\mathbf d,\mathbf B,\mathbf H\) in
\eqref{eq:sigma_onestep} denote the unweighted Gaussian cross-product
summaries.

\begin{algorithm}[h!]
\caption{One-Step Communication-Efficient Estimation for VCMMs (\(c=1\))}
\label{alg:onestep}
\begin{algorithmic}[1]
\STATE \textbf{Input:}
Data partitions
\(\{\mathcal D_s=(\mathbf y_s,\tilde{\mathbf X}_s,\mathbf Z_s)\}_{s=1}^K\),
pilot estimator
\(\widehat{\boldsymbol\theta}_0
=(\widehat{\tilde{\boldsymbol\beta}}_0^\top,
\widehat{\boldsymbol\alpha}_0^\top)^\top\) estimated from node \(1\), pilot
nuisance-parameter estimator
\(\widehat{\boldsymbol\eta}_0
=
(\widehat{\boldsymbol\Sigma}_{\alpha,0},
\widehat\sigma_{\epsilon,0}^2)\), penalty matrix
\(\mathbf P_{\lambda^*}\), and reference node \(1\).

\FOR{each node \(s=1,\ldots,K\)}
    \STATE Construct the pilot working response
    \(\mathbf y_s^{*(0)}\) and pilot working weight matrix
    \(\mathbf W_s^{(0)}\) using
    \((\widehat{\boldsymbol\theta}_0,\widehat{\boldsymbol\eta}_0)\).
    \STATE Compute the pilot-based local summary
    \(\boldsymbol\Gamma_s^{(0)}
    =(a_s^{(0)},\mathbf b_s^{(0)},\mathbf C_s^{(0)},
    \mathbf d_s^{(0)},\mathbf B_s^{(0)},\mathbf H_s^{(0)})\).
    \STATE Transmit \(\boldsymbol\Gamma_s^{(0)}\) once to the central server.
\ENDFOR

\STATE Aggregate the received summaries to form
\(\boldsymbol\Gamma^{(0)}
=
(a^{(0)},\mathbf b^{(0)},\mathbf C^{(0)},
\mathbf d^{(0)},\mathbf B^{(0)},\mathbf H^{(0)})\).

\STATE Compute the global gradient
\(\mathbf g(\widehat{\boldsymbol\theta}_0;\widehat{\boldsymbol\eta}_0)\)
using \eqref{eq:onestep_gradient}--\eqref{eq:onestep_gradient_components}.

\STATE Form the reference-node curvature approximation
\(\widetilde{\boldsymbol{\mathcal K}}\) using
\eqref{eq:onestep_reference_curvature}.

\STATE Compute the OS estimator
\(\widehat{\boldsymbol\theta}_{\mathrm{OS}}\) using
\eqref{eq:onestep_update}.

\STATE If the random-effect block in
\(\widetilde{\boldsymbol{\mathcal K}}\) is ill-conditioned, apply the
SVD-stabilized solve described in Appendix B.2.

\STATE Update
\(\widehat{\boldsymbol\eta}_{\mathrm{OS}}\) using
\eqref{eq:eta_onestep}.

\STATE \textbf{Output:}
\(\widehat{\boldsymbol\theta}_{\mathrm{OS}}\) and
\(\widehat{\boldsymbol\eta}_{\mathrm{OS}}\).
\end{algorithmic}
\end{algorithm}

In Algorithm~\ref{alg:onestep}, each node transmits its pilot-based working
summary only once. The aggregated summaries are used to compute the global
score at the pilot estimator, while the inverse curvature in the Newton-type
correction is provided by the reference-node approximation
\(\widetilde{\boldsymbol{\mathcal K}}^{-1}\). Therefore, the OS estimator is
not the full sufficient-statistics solution. It is a one-step
communication-efficient refinement from the pilot estimator toward the
full-data working objective, designed for the \(c=1\) regime.

We next establish the theoretical properties of the one-step estimator. Since
Algorithm~\ref{alg:onestep} uses one summary transmission and one Newton-type
refinement from a pilot estimator, the key question is whether this \(c=1\)
estimator can match the iterative sufficient-statistics estimator to first
order. The following theorem shows that this is the case when the pilot
estimator is sufficiently accurate and the curvature approximation is close to
the aggregated curvature matrix. Let
\(\widehat{\boldsymbol\theta}_{\mathrm{SS}}\) denote the full
sufficient-statistics estimator from Algorithm~\ref{alg:ss}, and let
\(\boldsymbol\theta^\star\) denote the population target. For reference, the
full aggregated curvature corresponding to the pilot working objective is
\begin{equation}
\label{eq:onestep_full_curvature}
\boldsymbol{\mathcal K}^{(0)}
=
\begin{pmatrix}
\mathbf C^{(0)}+\mathbf P_{\lambda^*}
&
\mathbf B^{(0)}
\\[0.35em]
\mathbf B^{(0)\top}
&
\mathbf H^{(0)}
+
\widehat{\boldsymbol\Sigma}_{\alpha,0}^{-1}
\end{pmatrix}.
\end{equation}

\begin{theorem}[First-order equivalence of the one-step estimator]
\label{thm:onestep}
Suppose Conditions~\textnormal{(B1)}--\textnormal{(B4)} in
Appendix C hold and the pilot estimator
\(\widehat{\boldsymbol\theta}_0\) is \(\sqrt N\)-consistent. If
\begin{equation}
\label{eq:curvature_condition}
\left\|
\widetilde{\boldsymbol{\mathcal K}}
-
\boldsymbol{\mathcal K}^{(0)}
\right\|_{\mathrm{op}}
=
o_p(N),
\end{equation}
then
\(\widehat{\boldsymbol\theta}_{\mathrm{OS}}
-
\widehat{\boldsymbol\theta}_{\mathrm{SS}}
=
o_p(N^{-1/2})\). Consequently, whenever
\(\widehat{\boldsymbol\theta}_{\mathrm{SS}}\) has limiting information
\(\boldsymbol{\mathcal I}\),
\(\sqrt N
(\widehat{\boldsymbol\theta}_{\mathrm{OS}}
-
\boldsymbol\theta^\star)
\overset{d}{\longrightarrow}
N(0,\boldsymbol{\mathcal I}^{-1})\).
\end{theorem}

\noindent
The proof is given in Appendix D of the Supplemental
Material. Theorem~\ref{thm:onestep} shows that the \(c=1\) estimator has the
same first-order behavior as the iterative SS estimator when the curvature
approximation is sufficiently accurate. Thus, a single communication-efficient
update can use the full-data score information encoded in the aggregated
pilot-based summaries, while relying on
\(\widetilde{\boldsymbol{\mathcal K}}^{-1}\) for a reference-node curvature
correction. When
\(\widetilde{\boldsymbol{\mathcal K}}=\boldsymbol{\mathcal K}^{(0)}\), the
condition in~\eqref{eq:curvature_condition} is automatically satisfied; the
theorem is most useful when \(\widetilde{\boldsymbol{\mathcal K}}\) is a
communication-reduced or SVD-stabilized approximation.

The first-order result establishes asymptotic equivalence between the one-step
estimator and the SS benchmark under a sufficiently accurate pilot and
curvature approximation. We next quantify this approximation in a growing
distributed-data regime, where the number of nodes \(K\) increases together
with the total sample size \(N\). In this regime, increasing \(K\) represents
adding independent distributed summaries rather than splitting a fixed dataset
into increasingly small pieces.

\begin{theorem}[Non-asymptotic deviation of the one-step estimator]
\label{thm:nonasymp}
Suppose Conditions~\textnormal{(B1)}--\textnormal{(B6)} in
Appendix C hold under the growing
distributed-data regime described above. Then there exist constants
\(c_1,c_2,c_3>0\), independent of \(K\), such that, for all \(u>0\),
\begin{equation}
\label{eq:theta_nonasymp}
\Pr\!\left(
\left\|
\widehat{\boldsymbol\theta}_{\mathrm{OS}}
-
\widehat{\boldsymbol\theta}_{\mathrm{SS}}
\right\|
\ge u
\right)
\le
c_1
\exp\!\left(
-\frac{c_2 Ku^2}{1+c_3u}
\right).
\end{equation}
Consequently, as \(K\to\infty\) and \(N\to\infty\),
\begin{equation}
\label{eq:theta_nonasymp_rate}
\left\|
\widehat{\boldsymbol\theta}_{\mathrm{OS}}
-
\widehat{\boldsymbol\theta}_{\mathrm{SS}}
\right\|
=
O_p\!\left(\sqrt{\frac{\log K}{K}}\right).
\end{equation}
\end{theorem}

\noindent
The proof is provided in Appendix~J of the Supplemental Material.
Theorem~\ref{thm:nonasymp} shows that the \(c=1\) one-step estimator
concentrates around the iterative SS benchmark as the distributed system grows.
The probability bound in~\eqref{eq:theta_nonasymp} implies
\eqref{eq:theta_nonasymp_rate} by taking
\(u=M\sqrt{\log K/K}\) for a sufficiently large constant \(M\). Hence, under
the stated regularity conditions, the discrepancy between the one-step
estimator and the SS estimator vanishes as \(K\) increases. This result should
be interpreted as a large distributed-data result, not as a claim that
splitting a fixed dataset into more nodes improves estimation.

Together with Theorem~\ref{thm:onestep}, this result justifies the one-step
estimator as a communication-efficient approximation to the iterative SS
estimator when only one communication round is allowed. The \(c>1\) regime uses
the SS estimator as an iterative sufficient-statistics solution and serves as
the full-data benchmark, while the \(c=1\) regime uses the OS estimator as a
single pilot-based refinement. Both methods avoid transmitting raw data and
full design matrices, and both handle large random-effect structures through
the summaries \(\mathbf d_s^{(r)}\), \(\mathbf B_s^{(r)}\), and
\(\mathbf H_s^{(r)}\), with \(r=0\) for OS.

\section{Numerical Studies}
\label{sec:numericalstudies}

We evaluate the proposed framework through grouped random-effect simulations
designed to assess both statistical accuracy and computational scalability. The
simulation is organized around two random-effect design structures. The first is
a dense uniform \(Z\) design, which creates a large dense random-effect structure. The second is an indicator \(Z\) design motivated by
origin--destination migration data, where each observation activates one origin
effect and one destination effect. Together, these settings assess whether the
proposed sufficient-statistics estimator (SS) and the communication-efficient
one-step estimator (OS) preserve full data accuracy while remaining feasible
when \(N\), \(q\), and the random-effect covariance structure become large. The
results provide empirical support for the likelihood-preserving property in
Theorem~\ref{thm:sufficient}, the first-order equivalence result in
Theorem~\ref{thm:onestep}, and the non-asymptotic deviation result in
Theorem~\ref{thm:nonasymp}.

All settings use the same varying-coefficient mean structure. The time-varying
coefficient is \(\beta_1(t)=\sin(2\pi t)\), approximated by cubic B-splines
with 20 internal knots placed uniformly on \((0,1)\) and boundary knots fixed
at 0 and 1. The spline penalty parameter is fixed at \(\lambda=1.0\), with no
cross-validation. The observations are split by batches: the last \(20\%\) of
batches are held out as an independent test set, and the remaining \(80\%\) are
used for training. The mean squared prediction error (MSPE) is computed on this
independent test set using the subject-specific predictor
$\widehat y_i
=
\tilde{\mathbf x}_i^\top \widehat{\tilde{\boldsymbol\beta}}
+
\mathbf z_i^\top \widehat{\boldsymbol\alpha}_{g_i},$
so both the estimated fixed-effect component and the estimated group-level
random-effect component are included in prediction. For runtime comparisons, we
record only the fitting time of each method; data generation, group-label
generation, spline-basis construction, train--test splitting, and test-set MSPE
evaluation are excluded. Unless otherwise stated, the data are distributed across \(K=20\)
computational nodes, which may be viewed as 20 parallel computer cores.

For the dense uniform \(Z\) design, the response is generated from
\[
    y_i
    =
    \beta_0
    +
    \sin(2\pi t_i)X_i
    +
    \mathbf z_i^\top\boldsymbol{\alpha}_{g_i}
    +
    \varepsilon_i,
    \qquad
    \varepsilon_i\sim N(0,\sigma_\varepsilon^2),
\]
where \(\beta_0=2\), \(t_i\sim \mathrm{Uniform}(0,1)\),
\(X_i\sim \mathrm{Uniform}(0,1)\), and \(\sigma_\varepsilon=0.25\). The group
labels \(g_i\) are assigned in a balanced way across \(G\) groups: observations
are divided equally and the group
labels are then randomly permuted. Thus, each group appears with approximately
equal frequency, while the assignment order remains random. Each entry of
\(\mathbf z_i\in\mathbb R^q\) is generated independently from
\(\mathrm{Uniform}(0,1)\), with no standardization.

For group \(g=1,\ldots,G\), let
\(\boldsymbol{\alpha}_g\in\mathbb R^q\) denote a group-specific random-effect
vector and write
$A=(\boldsymbol{\alpha}_1^\top,\ldots,\boldsymbol{\alpha}_G^\top)^\top
    \in \mathbb R^{G\times q}.$
The random-effect matrix is generated from
$A\sim MN_{G,q}(\mathbf 0,\Omega_G,\Sigma_q)$ and
$\mathrm{vec}(A)\sim N\{0,\Omega_G\otimes\Sigma_q\}$,
where \(\Omega_G\) captures between-group dependence and \(\Sigma_q\) captures
within-vector dependence. Both covariance matrices use AR(1) structures:
$(\Sigma_q)_{k\ell}
    =
    \sigma_\alpha^2\rho_{\mathrm{within}}^{|k-\ell|}$ and 
$(\Omega_G)_{gh}
    =
    \rho_{\mathrm{group}}^{|g-h|},$
with \(\sigma_\alpha=0.5\), \(\rho_{\mathrm{within}}=0.3\), and
\(\rho_{\mathrm{group}}=0.2\). This construction induces dependence both across
groups and across random-effect coordinates, producing a dense
\(Gq\times Gq\) covariance structure.

\begin{table}[h!]
\centering
\caption{MSPE and runtime comparisons under small and large sample sizes and
small and large random-effect dimensions. Values are reported as mean
(standard deviation). Runtime in panel~(b) is reported in seconds and refers
only to the fitting time of each method. The notation ``X'' indicates that the
method was skipped for the corresponding setting because of computational
infeasibility or statistical mismatch with the large 
random-effect structure.}
\label{tab:general_group_random_effect_sim}
\resizebox{\textwidth}{!}{
\begin{tabular}{lcccc}
\toprule
&
\shortstack{Uniform\\ \(N=10{,}000\)\\ \(q=2,\ G=150\)}
&
\shortstack{Uniform\\ \(N=100{,}000\)\\ \(q=300,\ G=150\)}
&
\shortstack{Uniform\\ \(N=1{,}000{,}000\)\\ \(q=300,\ G=150\)}
&
\shortstack{Indicator \(Z\)\\ \(N=1{,}000{,}000\)\\ \(q=300,\ G=150\)}
\\
\midrule
\multicolumn{5}{l}{\textbf{(a) MSPE}}\\
\midrule
OS
& \(0.1261\ (0.0043)\)
& \(0.2720\ (0.0036)\)
& \(0.1291\ (0.0008)\)
& \(0.1241\ (0.0010)\)
\\
SS
& \(0.1260\ (0.0043)\)
& \(0.2715\ (0.0035)\)
& \(0.1289\ (0.0006)\)
& \(0.1215\ (0.0004)\)
\\
tvReg
& \(0.2626\ (0.0245)\)
& X
& X
& X
\\
gamm4
& \(0.2621\ (0.0244)\)
& X
& X
& X
\\
lme4
& \(0.3271\ (0.0248)\)
& X
& X
& X
\\
\midrule
\multicolumn{5}{l}{\textbf{(b) Runtime}}\\
\midrule
OS
& \(0.01517\ (0.00852)\)
& \(0.34163\ (0.06947)\)
& \(0.52547\ (0.16324)\)
& \(0.28481\ (0.01480)\)
\\
SS
& \(0.12103\ (0.02827)\)
& \(9.92\ (1.50)\)
& \(15.76\ (1.50)\)
& \(0.37556\ (0.02706)\)
\\
tvReg
& \(104.54\ (35.02)\)
& X
& X
& X
\\
gamm4
& \(1.15\ (0.11)\)
& X
& X
& X
\\
lme4
& \(0.03760\ (0.01529)\)
& X
& X
& X
\\
\bottomrule
\end{tabular}
}
\end{table}

The dense uniform \(Z\) results in
Table~\ref{tab:general_group_random_effect_sim} show that OS and SS closely
agree across both small and large random-effect dimensions. In the
low-dimensional setting with \(N=10{,}000\) and \(q=2\), the two methods have
nearly identical MSPEs, \(0.1261\) for OS and \(0.1260\) for SS. The competing
methods are less accurate in this setting: tvReg and gamm4 have MSPEs around
\(0.26\), and lme4 has MSPE around \(0.33\). Although lme4 is computationally
fast in this low-dimensional setting, its statistical accuracy is weaker, while
tvReg is substantially slower because of its bandwidth-selection procedure.

When the random-effect dimension increases to \(q=300\), the advantage of the
proposed framework becomes more pronounced. At \(N=100{,}000\), OS and SS
remain nearly indistinguishable, with MSPEs \(0.2720\) and \(0.2715\),
respectively. At the million-sample scale, SS continues to match OS in MSPE
under the dense uniform design, with MSPEs \(0.1289\) and \(0.1291\),
respectively. These results support the convergence in Theorem~\ref{thm:sufficient} and its numerical stability in Appendix B.2, because the sufficient-statistics
updates remain accurate when the random-effect dimension is large. Although SS
is slower than OS in this dense setting because the sufficient summaries
involving \(\mathbf Z\) are more expensive to construct and manipulate, both
methods remain computationally feasible at \(N=1{,}000{,}000\) and \(q=300\).

The indicator \(Z\) design is used to mimic the origin--destination structure
in migration data. In this design, there are \(G\) groups, or areas, and
\(q=2G\), representing \(G\) origin random effects and \(G\) destination random
effects. Each row of \(\mathbf Z\in\mathbb R^{N\times q}\) contains exactly two
entries equal to one. Specifically, for each observation, an origin group
\(i\in\{1,\ldots,G\}\) and a destination group \(j\in\{1,\ldots,G\}\) are
sampled independently, with each group having equal probability of being
selected as an origin and equal probability of being selected as a destination.
The \(i\)-th column of \(\mathbf Z\) is set to one to activate the origin
effect, and the \((G+j)\)-th column is set to one to activate the destination
effect; all other entries are zero. Thus, the random-effect contribution for
that observation is the sum of one origin effect and one destination effect.
This construction produces a sparse random-effect design matrix and reflects
the push--pull structure used in the real-data migration analysis.

The final column of Table~\ref{tab:general_group_random_effect_sim} shows that
the proposed estimators remain accurate and computationally efficient under the
indicator \(Z\) design. At \(N=1{,}000{,}000\) and \(q=300\), SS achieves MSPE
\(0.1215\), compared with \(0.1241\) for OS. The runtimes are also close:
\(0.37556\) seconds for SS and \(0.28481\) seconds for OS. This efficiency is
expected because each row of \(\mathbf Z\) contains only two nonzero entries,
making the summaries \(\mathbf d=\mathbf Z^\top\mathbf y\),
\(\mathbf B=\tilde{\mathbf X}^\top\mathbf Z\), and
\(\mathbf H=\mathbf Z^\top\mathbf Z\) much cheaper to form and manipulate than
in the dense uniform design. These results are consistent with
Theorem~\ref{thm:onestep}, which explains why one-step refinement can retain
first-order efficiency relative to the iterative sufficient-statistics
benchmark.

Some competing packages are intentionally omitted from the large
random-effect settings in Table~\ref{tab:general_group_random_effect_sim}. The
tvReg package is omitted because its leave-one-out cross-validation procedure
for bandwidth selection has quadratic computational cost in \(N\), making it
computationally impractical for the large-\(N\) and large-\(q\) settings. The
gamm4 and lme4 packages are omitted because they are not designed to represent
the group-specific random-effect term
\(\mathbf z_i^\top\boldsymbol{\alpha}_{g_i}\) with a dense covariance structure
\(\Omega_G\otimes\Sigma_q\), and because they require full-data memory storage
and repeated matrix factorizations. Therefore, the skipped entries reflect
computational infeasibility and model incompatibility rather than missing
favorable comparisons.

Finally, we conduct a sensitivity analysis to examine the effect of the number
of distributed nodes on the discrepancy between OS and SS. Using the first
dense uniform \(Z\) setting in Table~\ref{tab:general_group_random_effect_sim},
we vary the number of nodes as \(K=5,10,20\). The discrepancy between the OS
and SS estimators decreases from \(0.0012\) to \(0.0005\) and then to
\(0.0001\). This monotone reduction is consistent with
Theorem~\ref{thm:nonasymp}, which states that the one-step estimator
concentrates around the SS benchmark as the number of independent distributed
summaries increases in a growing distributed-data regime. Thus, the sensitivity
analysis provides numerical evidence that the \(c=1\) estimator becomes closer
to the iterative SS estimator as the distributed system grows.

\section{Real Data Application: Large-Scale Migration Analysis}
\label{sec:realdata}

A major empirical challenge in studying modern migration is that the data are
large, spatially structured, and computationally demanding. We use the U.S.\
internal migration dataset compiled by \citet{HabansDouthat2024}, which
combines federal, state, and local migration records into a unified public
resource. The full collection contains approximately 1~TB of data and billions
of individual records from 2000 to 2020, including origin--destination
migration flows, socioeconomic covariates, and environmental indicators across
Commuting Zones (CZs). In this application, monthly migration
flows between \(154\) origin CZs and \(154\) destination CZs from 2000 to 2020
produce \(154\times154\times252=5{,}976{,}432\) CZ-to-CZ monthly observations,
or approximately six million records, when within-CZ flows are included.
Applying a VCMM to data of this size requires constructing spline-expanded
fixed-effect design matrices for time-varying coefficients and a random-effect
design matrix for origin and destination heterogeneity. Centralizing these
objects is computationally expensive and creates a communication bottleneck for
likelihood-based mixed-model estimation. These constraints motivate the
one-step communication-efficient estimator developed in
Section~\ref{sec:onestep_css}, which uses locally computed sufficient summaries
rather than transmitting raw observations or full design matrices.
In the implementation, the observations are randomly partitioned into
\(K=10\) approximately equal-sized distributed batches.
Each batch computes the local sufficient summaries
\((a_s,\mathbf b_s,\mathbf C_s,\mathbf d_s,\mathbf B_s,\mathbf H_s)\), which
are then aggregated to fit the model using the one-step communication-efficient
estimator (Algorithm 2).

To study temporal and spatial structure in U.S.\ migration flows, we fit the
following VCMM:
\begin{equation}
\label{eq:realmodel}
y_{ijt}
=
\beta_0(t)
+
\beta_1(t)\,\mathrm{DisasterDeclarations}_{ijt}
+
\mathbf Z_{ijt}^{\top}\boldsymbol{\alpha}
+
\varepsilon_{ijt},
\end{equation}
where \(y_{ijt}\) denotes the migration flow from origin CZ \(i\) to
destination CZ \(j\) at time \(t\). Specifically, \(y_{ijt}\) is
the monthly migration flow count for the CZ pair \((i,j)\) in month \(t\).
The function \(\beta_0(t)\) represents the time-varying baseline migration
level, while \(\beta_1(t)\) captures the time-varying association between
federally declared disasters and migration flows. The variable
\(\mathrm{DisasterDeclarations}_{ijt}\) is the number of declared
flood events in the destination CZ during month \(t\). The random-effect term
\(\mathbf Z_{ijt}^{\top}\boldsymbol{\alpha}\) captures spatial heterogeneity
through origin and destination effects. The time index \(t\)
corresponds to monthly observations from January 2000 through December 2020;
for spline estimation, the monthly index is represented as  $t = 1, \ldots, 252$.

With \(M=154\) CZs, we write
\[
\boldsymbol{\alpha}
=
\left(
\alpha^{O}_1,\ldots,\alpha^{O}_M,
\alpha^{D}_1,\ldots,\alpha^{D}_M
\right)^\top
\in\mathbb R^{2M}.
\]
Thus, the real-data model contains \(2M=308\) random-effect parameters.
The push effects are defined directly as the estimated origin
random effects \(\{\hat\alpha_i^O\}_{i=1}^M\), and the pull effects are defined
directly as the estimated destination random effects
\(\{\hat\alpha_j^D\}_{j=1}^M\). The origin-specific effect \(\alpha_i^O\) is
interpreted as outward migration pressure, or a push effect, and the
destination-specific effect \(\alpha_j^D\) is interpreted as destination
attractiveness, or a pull effect. The design vector
\(\mathbf Z_{ijt}\in\mathbb R^{308}\) contains exactly two nonzero entries:
one for the origin CZ and one for the destination CZ. Hence, $\mathbf Z_{ijt}^{\top}\boldsymbol{\alpha}
=
\alpha_i^O+\alpha_j^D.$

The covariance matrix of \(\boldsymbol{\alpha}\) also has a direct migration
interpretation. We write it as a \(2\times2\) block covariance matrix,
\[
\boldsymbol{\Sigma}_\alpha
=
\begin{pmatrix}
\boldsymbol{\Sigma}_{OO} & \boldsymbol{\Sigma}_{OD} \\
\boldsymbol{\Sigma}_{DO} & \boldsymbol{\Sigma}_{DD}
\end{pmatrix},
\qquad
\boldsymbol{\Sigma}_{OO},\boldsymbol{\Sigma}_{OD},
\boldsymbol{\Sigma}_{DO},\boldsymbol{\Sigma}_{DD}
\in\mathbb R^{154\times154}.
\]
The block \(\boldsymbol{\Sigma}_{OO}\) describes dependence among origin push
effects, \(\boldsymbol{\Sigma}_{DD}\) describes dependence among destination
pull effects, and the off-diagonal blocks
\(\boldsymbol{\Sigma}_{OD}\) and \(\boldsymbol{\Sigma}_{DO}\) describe
dependence between push and pull effects.  This \(308\times308\) random-effect covariance matrix is left unstructured,
rather than being constrained to a separable, block-diagonal, or prespecified
dependence structure. Thus, it is richer than the random-effect structures commonly used in applied migration models.

Although each row of \(\mathbf Z_{ijt}\) is sparse, the full random-effect
design matrix is constructed over millions of origin--destination--time
observations. Direct centralized estimation is therefore computationally
demanding. Model~\eqref{eq:realmodel} provides a structured decomposition of
migration flows into time-varying fixed effects and large-scale spatial random
effects, while the proposed sufficient-statistics framework avoids direct
transmission or storage of the full design matrices. The temporal evolution of
\(\beta_0(t)\) and \(\beta_1(t)\) is examined in
Section~\ref{sec:tvcoef}, and the estimated push and pull effects are examined
in Section~\ref{sec:pullpush}.

\subsection{Estimated Time-Varying Coefficients}
\label{sec:tvcoef}

This subsection summarizes the estimated time-varying coefficient functions in
model~\eqref{eq:realmodel}. The temporal functions \(\beta_0(t)\) and
\(\beta_1(t)\) are represented using cubic B-spline basis functions with 58
knots. The knots are placed over the monthly time domain from 2000
to 2020, and the smoothing parameter is
selected by cross-validation. This flexible
representation allows the fitted model to capture nonlinear and time-varying
migration patterns while maintaining a stable spline-based estimation
structure. Estimation is carried out using the one-step communication-efficient
estimator, which aggregates local sufficient summaries across distributed data
batches.

\begin{figure}[htbp]
\centering
\begin{minipage}{0.48\textwidth}
\centering
\includegraphics[width=\textwidth]{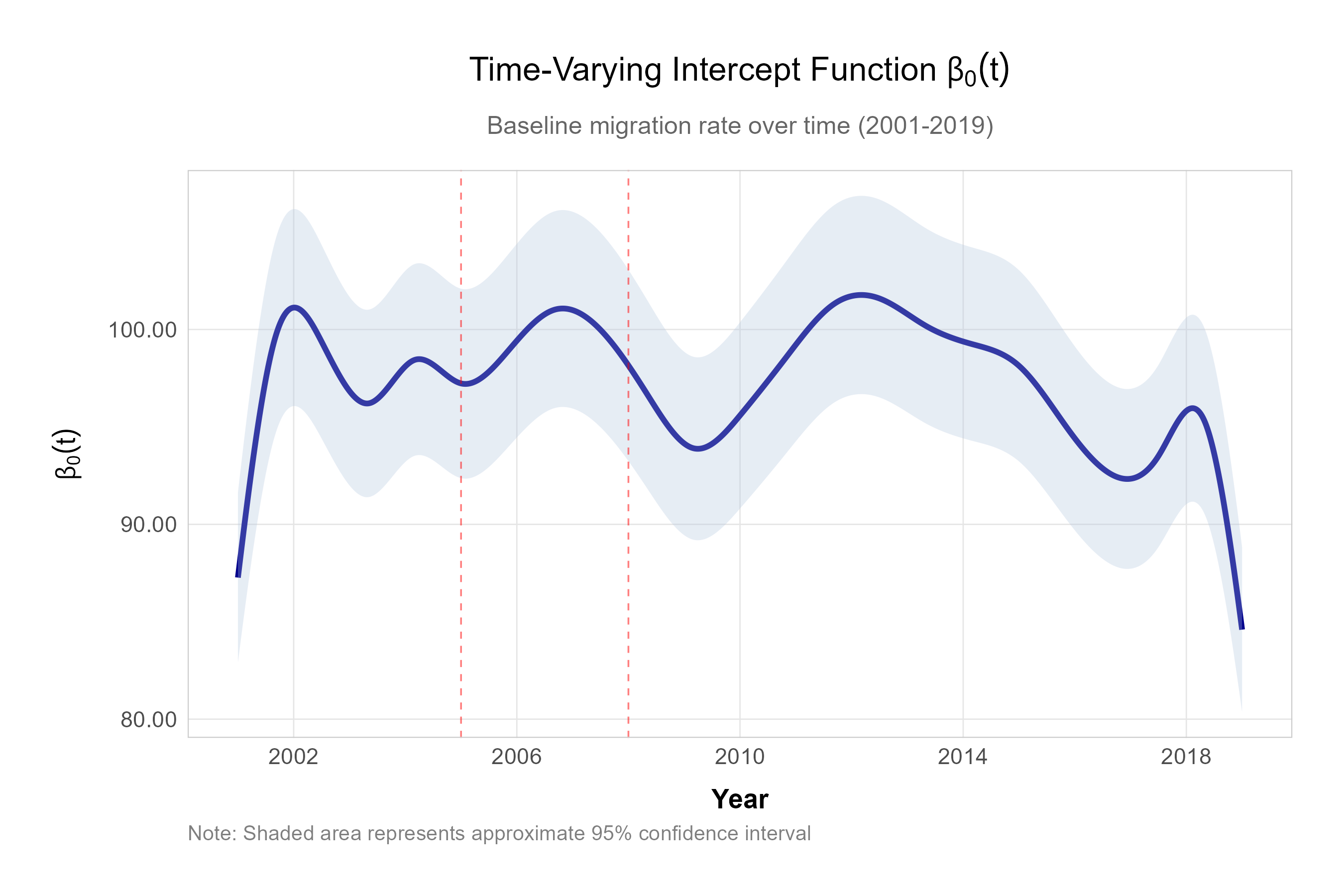}
\end{minipage}
\hfill
\begin{minipage}{0.48\textwidth}
\centering
\includegraphics[width=\textwidth]{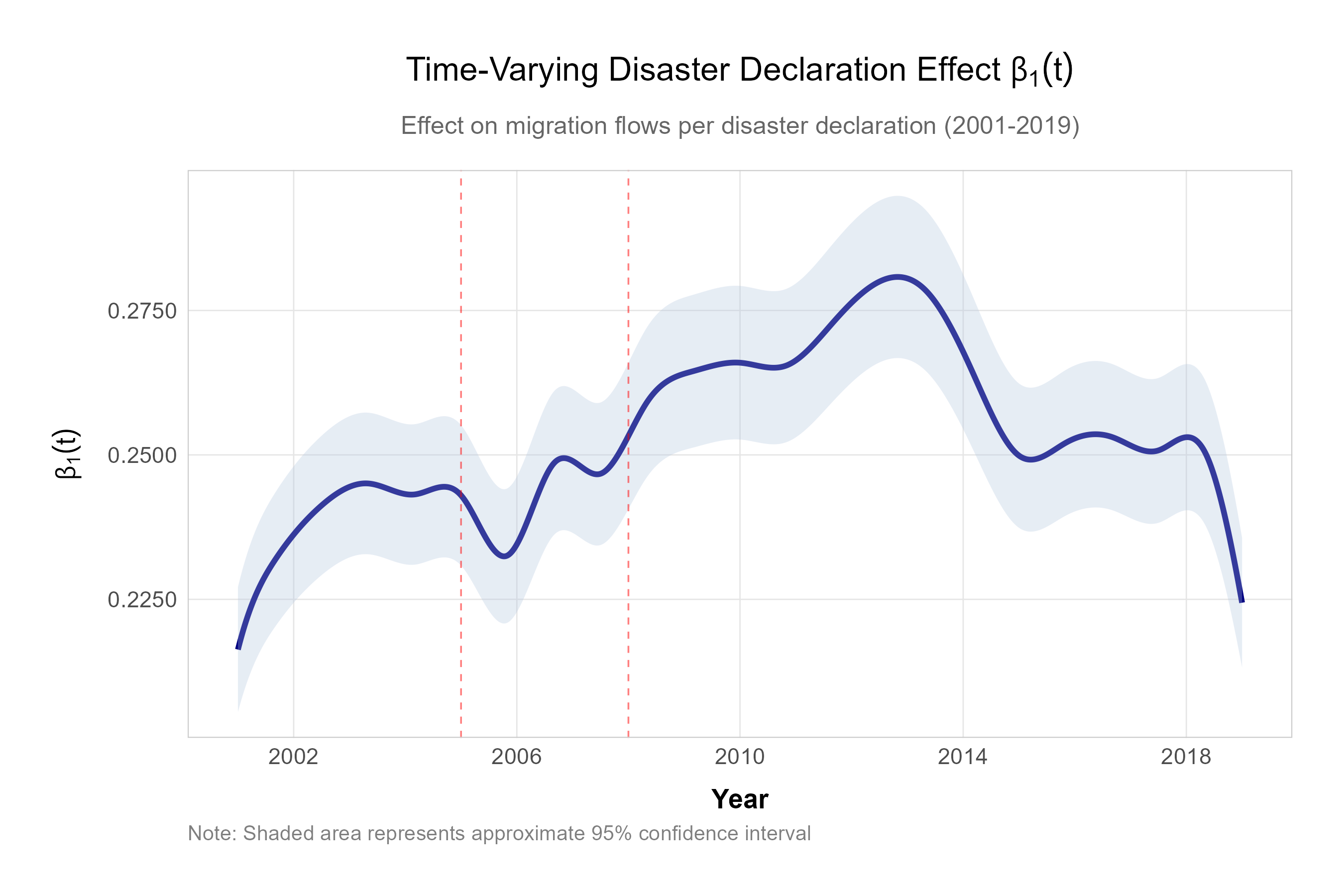}
\end{minipage}
\caption{Estimated time-varying coefficient functions: (Left) baseline
migration function \(\beta_0(t)\) from 2000 to 2020; (Right)
disaster-declaration effect \(\beta_1(t)\). Shaded regions denote
pointwise 95\% confidence intervals.}
\label{fig:time_varying_functions}
\end{figure}

Figure~\ref{fig:time_varying_functions} displays the estimated baseline
function \(\beta_0(t)\) and disaster-declaration effect \(\beta_1(t)\). The
baseline function \(\beta_0(t)\) shows clear temporal variation over the study
period. It increases in the early 2000s, rises again near 2010, reaches another
local maximum around 2015, and then declines toward the end of the sample
period. This pattern suggests that the baseline level of migration is not
constant over time. The narrow 95\% confidence bands indicate that this
temporal pattern is estimated precisely under the fitted model. 
The shaded bands are pointwise 95\% confidence intervals computed from the estimated information matrix in Theorem 3.1.

The disaster-declaration coefficient \(\beta_1(t)\) also changes over time. It
is lowest around 2005--2006, a period that includes Hurricane Katrina and
substantial Gulf Coast displacement, and then increases through the early
2010s. After approximately 2015, the estimated effect declines. This pattern
indicates that the association between federally declared disasters and
migration flows is time-varying rather than constant. These estimated
time-varying coefficients provide the fixed-effect component of the analysis,
while the origin and destination random effects capture additional spatial
heterogeneity.

\subsection{Push and Pull Effects}
\label{sec:pullpush}

The random-effect component in model~\eqref{eq:realmodel} decomposes spatial
heterogeneity into origin-specific and destination-specific effects.
Throughout this section, the push effects refer directly to the
estimated origin random effects \(\hat\alpha_i^O\), and the pull effects refer
directly to the estimated destination random effects \(\hat\alpha_j^D\).
The origin-specific effects \(\{\alpha_i^O\}\) measure outward migration
pressure, or push effects, while the destination-specific effects
\(\{\alpha_j^D\}\) measure destination attractiveness, or pull effects. This
decomposition is useful because each migration flow depends jointly on the push
effect of the origin and the pull effect of the destination. The complete
CZ-level random-effect estimates are reported in
Appendix~G.1.

The estimated random effects show substantial spatial heterogeneity across the
154 CZs. Destination pull effects range from approximately \(-99.097\) to
\(354.502\), while origin push effects range from approximately \(-136.363\) to
\(106.533\). The empirical correlation between the estimated pull and push
effects is approximately \(0.555\), suggesting that CZs with stronger
destination attractiveness also tend to have stronger outward mobility after
adjustment for the time-varying fixed effects. This pattern is consistent with
high-mobility regional systems in which population inflows and outflows can both
be large.

Several major metropolitan areas have large positive destination effects.
Houston has the largest pull effect \((354.502)\) and a positive push effect
\((16.227)\). New Orleans has the second-largest pull effect \((238.703)\) and
the largest push effect \((106.533)\). Other high-pull CZs include Tampa
\((166.549)\), Dallas \((152.425)\), San Antonio \((120.554)\), Port St. Lucie
\((112.513)\), Miami \((96.117)\), Baton Rouge \((90.725)\), Orlando
\((78.987)\), and Lafayette \((57.471)\). These estimates indicate that several
large or high-growth metropolitan areas remain strong migration destinations
after accounting for temporal baseline migration and disaster-declaration
effects. In contrast, Guymon \((-99.097)\), Seymour \((-90.983)\), Clovis
\((-88.194)\), Brady \((-80.894)\), Haskell \((-78.597)\), and Corsicana
\((-76.433)\) have strongly negative pull effects.

The push effects provide complementary information about outward migration
pressure. New Orleans is the most distinctive CZ, with the largest positive
push effect \((106.533)\) and the second-largest pull effect \((238.703)\).
This suggests that New Orleans is both a strong destination and a high-mobility
origin after adjustment for the fixed effects. Houston also combines the largest
pull effect with a positive push effect, suggesting a dynamic metropolitan
migration system. At the lower end, Pearsall has the smallest push effect
\((-136.363)\), followed by Childress \((-114.116)\), Alpine \((-102.734)\),
New Albany \((-99.135)\), and Demopolis \((-98.002)\).

Within Louisiana, the estimated random effects show clear regional differences.
New Orleans has both a large pull effect and the largest push effect. Baton
Rouge has a strong destination effect \((90.725)\) and a modestly negative push
effect \((-14.674)\), suggesting a more stabilizing regional role. Lafayette
has a positive pull effect \((57.471)\) and a lower push effect
\((-40.558)\). Shreveport \((16.592,-41.758)\), Monroe
\((10.612,-78.805)\), Houma \((9.092,-51.282)\), and Lake Charles
\((-2.640,-47.797)\) show more moderate or weaker destination effects. These
results suggest that Louisiana contains both high-mobility recovery centers and
more stable regional destinations. Compared with traditional gravity models
\citep{porojan2001trade}, the VCMM random-effect formulation provides a direct
decomposition of spatially varying push and pull effects after adjustment for
time-varying baseline migration and disaster-declaration effects.

\begin{figure}[h!]
\centering
\includegraphics[width=\textwidth]{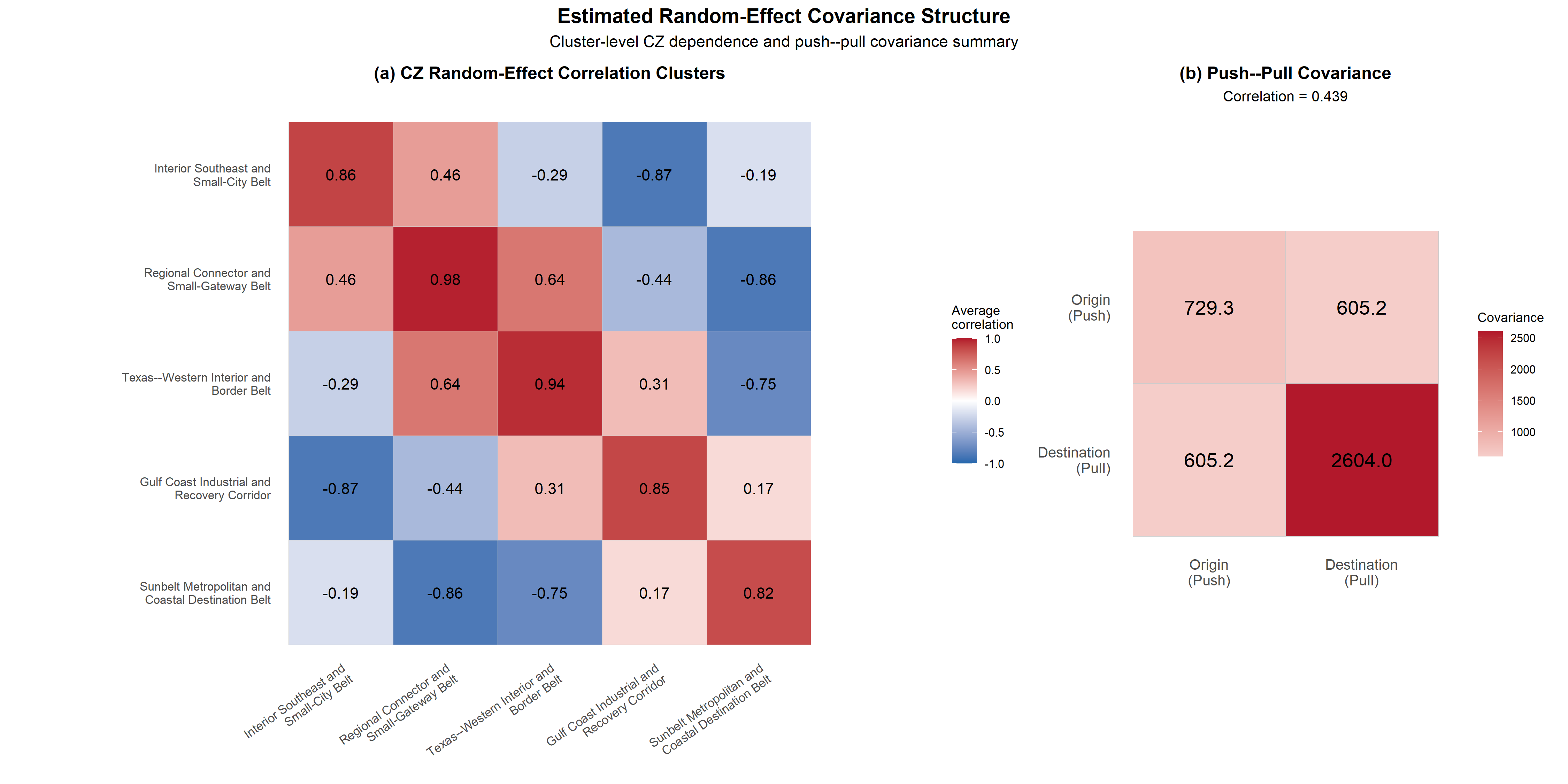}
\caption{Estimated random-effect covariance structure. Panel~(a) reports a
five-cluster summary of the estimated CZ random-effect correlation matrix. Each
cell gives the average correlation between two CZ groups computed from the full
\(154\times154\) random-effect correlation matrix; within-cluster averages
exclude diagonal self-correlations. Panel~(b) reports the estimated \(2\times2\)
push--pull random-effect covariance matrix. The diagonal entries are the
variances of the origin, or push, and destination, or pull, effects; the
off-diagonal entries give their covariance.}
\label{fig:covariance_summary_main}
\end{figure}

Figure~\ref{fig:covariance_summary_main} summarizes the estimated random-effect
covariance structure. Panel~(a) reports a five-cluster summary of the estimated
CZ-level random-effect correlation matrix. The five CZ groups are
obtained by hierarchical clustering of the estimated \(154\times154\) CZ
random-effect correlation matrix. Specifically, we first convert the estimated
covariance matrix to a correlation matrix and then cluster CZs using the
correlation-profile distance
\(1-\mathrm{cor}(\boldsymbol r_i,\boldsymbol r_j)\), where
\(\boldsymbol r_i\) and \(\boldsymbol r_j\) are the correlation profiles of CZs
\(i\) and \(j\). Average-linkage hierarchical clustering is used, and the tree
is cut into five clusters. The cluster names are assigned post hoc to summarize
the dominant geographic and migration-system features of the CZs in each
cluster. Positive diagonal blocks indicate stronger within-cluster dependence,
while negative off-diagonal entries indicate contrasting migration behavior
between some regional groups. Panel~(b) reports the estimated \(2\times2\)
push--pull covariance matrix. The destination variance is larger than the
origin variance \((2603.998\) versus \(729.284)\), suggesting that destination
attractiveness varies more strongly across CZs than outward migration pressure.
The positive origin--destination covariance \((605.170)\), with implied
push--pull correlation \(0.439\), indicates positive dependence between the two
random-effect components. These results show that the estimated covariance
structure captures interpretable dependence among regional migration systems,
rather than serving only as a numerical component of the mixed model.
Additional covariance summaries and interpretation are provided in
Appendix~G.2.

\begin{figure}[htbp]
\centering
\includegraphics[width=\textwidth]{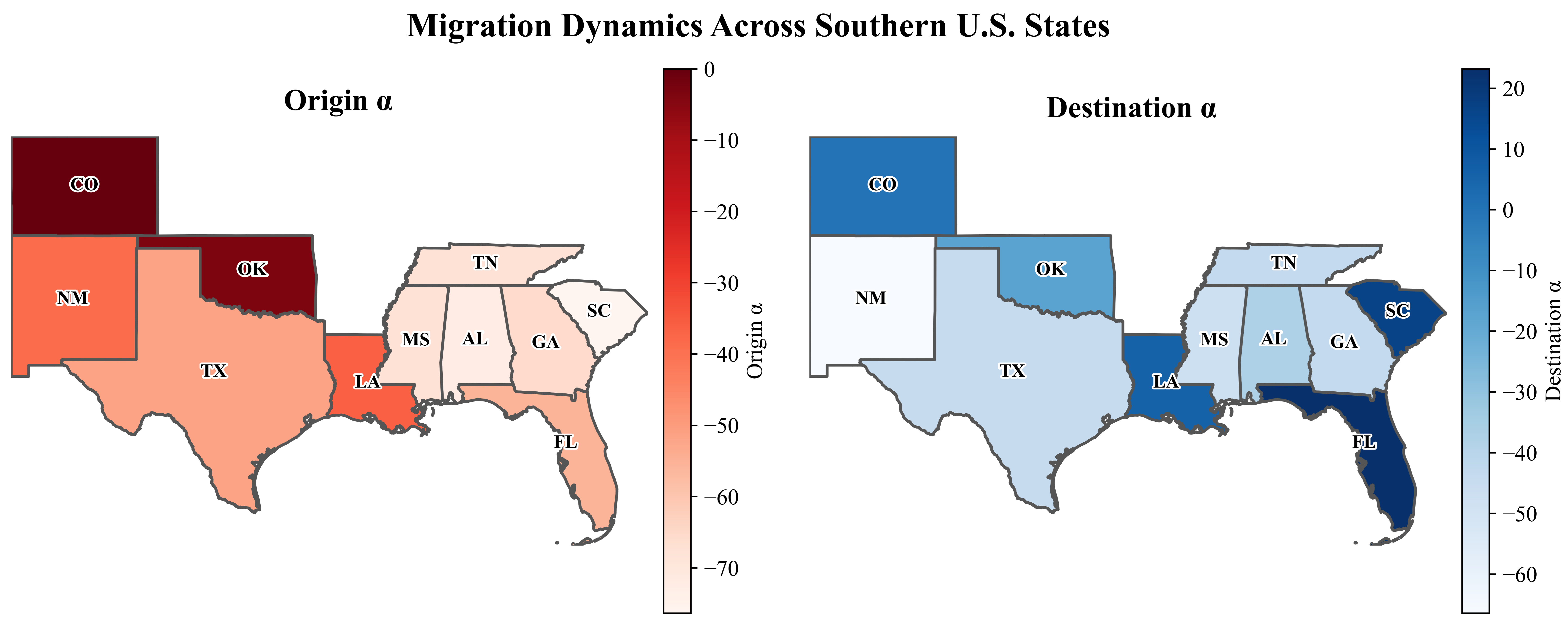}
\caption{Southern-state migration effects based on estimated random effects. The left panel shows the estimated origin effect \(\widehat{\alpha}^{O}\),
representing emigration pressure or push effects, and the right panel shows the estimated destination effect \(\widehat{\alpha}^{D}\), representing immigration
attractiveness or pull effects.}
\label{fig:southern_states_migration}
\end{figure}

Figure~\ref{fig:southern_states_migration} displays the geographic structure of
the estimated random effects across southern and coastal states. The maps show
spatial clustering in the estimated origin and destination random effects,
especially along the Gulf Coast and the southeastern seaboard. Coastal and
metropolitan regions tend to have stronger destination effects, while some
inland regions show weaker migration effects. The maps display
the estimated random effects \(\hat{\boldsymbol{\alpha}}\) directly, rather
than differences between push and pull effects. These results suggest that
environmental exposure alone does not determine migration outcomes; some exposed
regions remain attractive destinations, while other regions show weaker
migration performance.

Overall, the real-data analysis shows that the proposed VCMM separates three
sources of variation: nonlinear temporal baseline migration, time-varying
disaster-declaration effects, and spatially heterogeneous push--pull random
effects. The one-step communication-efficient estimator enables this analysis
without transmitting raw data or full design matrices across nodes.
Appendix~G.1 reports the full CZ-level push
and pull estimates, and Appendix~G.2 provides
additional covariance analysis.

\section{Summary and Future Directions}\label{sec:conclusion}

This paper develops a unified framework for varying coefficient mixed models that operates effectively in both centralized and distributed large-scale settings. In the unconstrained regime, we show that penalized spline estimators admit an exact Bayesian hierarchical representation, yielding a full-information likelihood with classical convergence rates and optimality. This formulation reveals a fixed, low-dimensional set of sufficient statistics that fully encode each node’s likelihood contribution, enabling communication-efficient estimation when raw data or large spline design matrices cannot be transferred. Aggregating these summaries produces a surrogate likelihood supporting a one-step estimator that attains the same first-order efficiency as the centralized maximum likelihood estimator, while SVD-enhanced implementation ensures numerical stability under large-dimensional spline bases and correlated random effects. Together, these components form a coherent and scalable toolkit for VCMM estimation across a range of communication environments.

The practical value of the method is demonstrated through a large-scale analysis of U.S. internal migration linked to federally declared disasters. To our knowledge, this is the first application to fit a full VCMM with millions of
origin--destination flows under communication constraints. The fitted model reveals substantial spatial heterogeneity in migration push--pull dynamics and
captures long-run macroeconomic and disaster-related temporal patterns, demonstrating that communication-aware VCMM estimation can achieve production-scale scalability while preserving statistical efficiency and uncertainty quantification. Future work will extend this framework to real-time inference based on streaming sufficient statistics, enabling one-step parameter updates as new data arrive and supporting dynamic monitoring of large-scale spatiotemporal systems.



\bibliographystyle{apalike}
\bibliography{references}

\end{document}